\documentstyle[aps,twocolumn]{revtex}
\begin{document}

\newcommand{\be}{\begin{equation}}
\newcommand{\ee}{\end{equation}}
\newcommand{\bea}{\begin{eqnarray}}
\newcommand{\eea}{\end{eqnarray}}
\newcommand{\nn}{\nonumber}

\title{ On the Infrared Limit of Unconstrained $SU(2)$ Yang-Mills Theory }
\author{A. M. Khvedelidze \ $^a$
\thanks{Permanent address: Tbilisi Mathematical Institute,
380093, Tbilisi, Georgia} and
\,\, H.-P. Pavel\ $^b$ }
\address{$^a$ Bogoliubov Theoretical Laboratory, Joint Institute for Nuclear
Research, Dubna, Russia}
\address{$^b$ Fachbereich Physik der Universit\"at Rostock,
              D-18051 Rostock, Germany}
\date{\today}
\maketitle

\begin{abstract}

The variables appropriate for the infrared limit
of unconstrained $SU(2)$ Yang-Mills field theory are obtained
in the  Hamiltonian formalism.
It is shown how in the infrared limit an effective nonlinear sigma model 
type Lagrangian can be derived which out of the six physical fields involves 
only one of three scalar fields and two rotational fields summarized 
in a unit vector. Its possible relation to the effective Lagrangian proposed 
recently by Faddeev and Niemi is discussed.
 
\end{abstract}


\bigskip


\bigskip

The conventional perturbative treatment of gauge theories
works successfully for the description of
high energy phenomena,  but fails in applications in the infrared region.
Several different approaches \cite{GoldJack}-\cite{KJohnson} have been
proposed for the nonperturbative reduction of gauge theories
to the equivalent unconstrained system.
The guideline of these investigations has been the search for a
representation of the gauge invariant
variables which  are adapted to the study of the low energy phase
of Yang-Mills theory.
An alternative and very interesting approach has been proposed
very recently by \cite{FadNiem} where a topological soliton model
with features relevant for the low energy region is argued to be extendable
to full $SU(2)$ Yang-Mills theory.
We shall discuss in this work how an effective low energy theory
can be obtained directly from the unconstrained sytem.

In previous work \cite{KP} we obtained the unconstrained system equivalent
to the degenerate $SU(2)$ Yang-Mills theory
following the method of Hamiltonian reduction \\ (\cite{GKP,GKMP}
and references therein)
in the framework of the  Dirac constraint formalism \cite{DiracL}.
It has been shown that $SU(2)$ Yang-Mills theory
can be reduced to the corresponding unconstrained system
describing the dynamics of a
positive definite symmetric $3\times 3$ matrix \( Q^\ast \).
In this letter we separate the six physical dynamical field variables
into three rotation invariant and three rotational ones.
We shall obtain an effective low energy theory involving only two of the 
three rotational fields, summarized in a unit vector, and one of the tree 
scalar fields, and shall discuss its possible relation to the effective 
soliton Lagrangian proposed recently in \cite{FadNiem}.

As derived in \cite{KP} the dynamics of the physical variables of $SU(2)$
Yang-Mills theory can be described by the nonlocal Hamiltonian
\be
\label{eq:uncYME}
H =\frac{1}{2} \int d^3{x}
\biggl[
\mbox{Tr}(P^\ast)^2 + \mbox{Tr}(B^2(Q^\ast))
+ {1\over 2}{\vec E}^2(Q^\ast,P^\ast)
\biggr]
\ee
in terms of the unconstraint canonical pairs, the positive definite symmetric
$3\times 3$ matrices \(Q^\ast\) and \(P^\ast\).
The first term is the conventional quadratic ``kinetic'' part,
the second the trace of the square of the non-Abelian magnetic field
\be
B_{sk}(Q^{\ast})= \epsilon_{klm}
(\partial_l Q^{\ast}_{sm} +
{g\over 2}\epsilon_{sbc} \, Q^{\ast}_{bl}Q^{\ast}_{cm})~.
\ee
The third term in the Hamiltonian is the square of the vector ${\vec E}$
given as solution of the differential equation
\be
\label{vecE}
\left[\gamma_{ik}(Q^{\ast})-{1\over g}\epsilon_{ikl}\partial_l\right]E_{k}
= {\cal S}_i
\ee
with $\gamma_{ik}:= Q^{\ast}_{ik}-\delta_{ik}\mbox{Tr}(Q^{\ast})$ and the
source term
\be
\label{eq:spin}
{\cal S}_k (x) := \epsilon_{klm}\left(P^\ast Q^{\ast}\right)_{lm} -
{1\over g} \partial_l P_{kl}^\ast~,
\ee
which coincides with the spin density part of the Noetherian angular momentum
up to divergence terms.

The solution ${\vec E}$ of the differential equation (\ref{vecE})
can be expanded in $1/g$. The zeroth order term is
\be
E^{(0)}_{s}=\gamma^{-1}_{sk}\epsilon_{klm}\left(P^\ast Q^{\ast}\right)_{lm}~,
\ee
and the first order term is determined via
\be
E^{(1)}_{s} := {1\over g}
\gamma^{-1}_{sl}\left[(\mbox{rot}\ {\vec {E}}^{(0)})_l
-\partial_k P^\ast_{kl}\right]
\ee
from the corresponding zeroth order term.
The higher terms are obtained via the simple recurrence relations
\be
\label{vecE2}
E^{(n+1)}_{s} := {1\over g}
\gamma^{-1}_{sl}(\mbox{rot}\ {\vec {E}}^{\ (n)})_l~.
\ee

Whereas the gauge fields transform as vectors under spatial rotations,
the unconstrained fields $Q^\ast$ and $P^\ast$
transform as second rank tensors under spatial rotations.\footnote{
Note that for a complete analysis it is necessary to investigate
the transformation properties of the field $Q^\ast$ under the whole
Poincar\'e group. We shall limit ourselves here to the isolation
of the scalars under spatial rotations and
treat $Q^\ast$ in terms of ``nonrelativistic  spin 0 and spin 2
fields'' in accordance with the conclusions obtained in the work
\cite{Faddeev79}.}
In order to separate the three fields which are invariant under spatial
rotations from the three rotational degrees of freedom we perform the
following main axis transformation of the original positive definite symmetric
$3\times 3$ matrix field $Q^\ast(x)$
\be
\label{mainaxis}
Q^\ast\left(\chi,\phi\right) =
R^T(\chi(x)){\cal D}\left(\phi(x)\right) R(\chi(x)),
\ee
with the orthogonal matrix $R(\chi)$
and the positive definite diagonal matrix
\be
{\cal D}\left(\phi\right)
:=\mbox{diag}\left(\phi_1, \phi_2,\phi_3\right),\ \ \ \phi_i > 0
\ \ ( i=1,2,3),
\ee
the unconstrained Hamiltonian can be written in the form
\be
\label{eq:unchdia}
H \, = \,
\frac{1}{2} \int d^3x
\left(\sum_{i=1}^3\pi_i^2 +
 {1\over 2}\sum_{i=1}^3{\cal P}_i^2 +
{1\over 2}{\vec{\cal E}}^{\ 2} + V  \right)~.
\ee
Here the fields $\pi_i$ are the canonically conjugate momenta to the diagonal
fields $\phi_i$ and
\be
{\cal P}_{i}(x):=\frac{\xi_i(x)}{\phi_j(x) - \phi_k(x)},
 \,\,
(cycl.\,\,\, perm.\,\,\, i\not=j\not= k )
\ee
with the $SO(3)$ left-invariant Killing vectors
\be
\xi_k(x): ={\cal M}(\theta, \psi)_{kl} p_{\chi_l}~,
\ee
where in terms of the Euler angles
\be \label{eq:MCmatr}
{\cal M}(\theta, \psi): =
\left (
\begin{array}{ccc}
\sin\psi/\sin\theta,     & \cos\psi,     & - \sin\psi\cot\theta     \\
-\cos\psi/\sin\theta,    & \sin\psi ,    &  \cos\psi\cot\theta      \\
            0,                  &   0 ,         &     1
\end{array}
\right ).
\ee
The electric field vector ${\vec {\cal E}}$
is given by an expansion in $1/g$
with the zeroth order term
\be
{\cal E}^{(0)}_{i} := -\frac{\xi_i}{\phi_j+ \phi_k}
 \,\,\,\,\,
(cycl.\,\,\,\, permut. \,\,\, i\not=j\not= k )~,
\ee
the first order term given from ${\cal E}^{(0)}$ via
\be
{\cal E}^{(1)}_{i} := - {1\over g}\frac{1}{\phi_j+\phi_k}
\left[\left((\nabla_{X_j}{\cal E}^{(0)})_k-(\nabla_{X_k}{\cal E}^{(0)})_j
\right)+\Xi_i\right]\ee
with cyclic permutations of $ i\not=j\not= k $ and
the higher order terms of the expansion determined via the recurrence
relations
\be
{\cal E}^{(n+1)}_{i} := - {1\over g}\frac{1}{\phi_j+\phi_k}
\left((\nabla_{X_j}{\cal E}^{(n)})_k-(\nabla_{X_k}{\cal E}^{(n)})_j\right)~.
\ee
Here the components of the covariant derivatives $\nabla_{X_k}$
along the vector fields $X_k:=R_{ki}\partial_i$
\be
(\nabla_{X_k}\vec{\cal E})_b :=
X_k {\cal E}_b + \Gamma^d_{{\ }kb}{\cal E}_d
\ee
are determined by the connection
\be
\label{Connection}
\Gamma^b_{{\ }i a} :=
      \left( R X_i R^T\right)_{ab}
     =- \epsilon_{abs} ({\cal M}^{-1})_{sk} X_i \chi_k~.
\ee
The ``source'' terms $\Xi_k$ are given as
\bea
\Xi_1&=& \Gamma^1_{{\ }2 2}(\pi_1-\pi_2)+{1\over 2}X_1\pi_1
       -\Gamma^2_{{\ }2 3}{\cal P}_2-\Gamma^1_{{\ }2 3}{\cal P}_1\nonumber\\
     &&  -2\Gamma^1_{{\ }1 2}{\cal P}_3+X_2{\cal P}_3
       \ +\ (2 \leftrightarrow 3)
\eea
and its cyclic permutaions $\Xi_2$ and $\Xi_3$.

The magnetic part $V$ of the potential is
\be
\label{Vinhom1}
V[\phi,\chi] = \sum_{i=1}^3 V_i[\phi,\chi]
\ee
with
\bea
\label{Vinhom2}
V_1[\phi,\chi] &=& \left(\Gamma^1_{{\ }12}(\phi_2-\phi_1)
                              -X_2 \phi_1\right)^2 \nonumber\\
                  & +& \left(\Gamma_{{\ }13}^1(\phi_3-\phi_1)
                      -X_3 \phi_1\right)^2 \nonumber\\
                  & +&\left(\Gamma_{{\ }23}^1\phi_3
                +\Gamma_{{\ }32}^1\phi_2-g\phi_2 \phi_3\right)^2
\eea
and its cyclic permutations .

In the strong coupling limit the expression (\ref{eq:unchdia}) for the
unconstrained Hamiltonian reduces to
\footnote{For spatially constant fields the integrand of this expression
reduces to the Hamiltonian of $SU(2)$ Yang-Mills mechanics considered
in previous work \cite{GKMP}.}
\be
\label{eq:strongh}
H_{S} = 
\frac{1}{2}\int d^3x \left(\sum_{i=1}^3 \pi_i^2
 + \sum_{cycl.}\xi^{2}_i\frac{\phi_j^2+\phi_k^2 }{(\phi_j^2-\phi_k^2)^2}
+ V[\phi,\chi] \right)
\ee
For the further investigation of the low energy properties of $SU(2)$
field theory a thorough understanding of the properties of the term in
(\ref{Vinhom1}) containing no derivatives
\be
\label{Vhomo}
V_{\rm hom}[\phi_i] = g^2[\phi_1^2\phi_2^2+\phi_2^2\phi_3^2+\phi_3^2\phi_1^2]
\ee
is crucial. The classical absolute minima of energy correspond to
vanishing of the positive definite kinetic term
in the Hamiltonian (\ref{eq:strongh}). The stationary points of the
potential term (\ref{Vhomo}) are
\be
\phi_1=\phi_2=0\ ,\ \ \ \phi_3\  {\rm arbitrary}
\ee
and its cyclic permutations.
Analysing the second order derivatives of the potential
at the stationarity points one can conclude that
they form a continous line of degenerate absolute minima at zero energy.
In other words the potential has a ``valley'' of zero energy minima
along the line $\phi_1=\phi_2=0$. They are the unconstrained analogs of the
toron solutions \cite{Luescher} representing constant
Abelian field configurations with vanishing magnetic field
in the strong coupling limit.
The special point $\phi_1=\phi_2=\phi_3=0$ corresponds to the ordinary
perturbative minimum.

For the investigation of the configurations of higher energy
it is necessary to include the rotational term in (\ref{eq:strongh}).
Since the singular points of the rotational term just correspond to the
absolute minima of the potential there will a competition between an attractive
and a repulsive force. At the balance point we will have a local minimum
corresponding to a classical configuration with higher energy.

The above representation (\ref{mainaxis}) in terms of scalar and 
rotational fields gives us furthermore the possibility to analyse 
the wellknown exact, but nonnormalizable,
zero energy groundstate wave functional 
of $SU(2)$ gluodynamics \cite{Loos} in the strong coupling limit.
For the corresponding unconstrained Hamiltonian (\ref{eq:uncYME})
it has been discussed in \cite{KP} and has the form
\be
\label{PsiQ}
\Psi[Q^\ast] = \exp{\left(- 8\pi^2 W[Q^\ast]\right)}
\ee
with the winding number
functional \cite{Jackiw} $W[Q^\ast]: = \int d^3x\  K_0(x)$ in terms of
the zero component of the Chern-Simons vector \cite{Deser},
$K_0(Q^\ast): = -(16\pi^2)^{-1} \epsilon^{ijk}
\mbox{Tr}\left(F_{ij} Q^\ast_k -\frac{2}{3}gQ^\ast_i Q^\ast_j Q^\ast_k
\right)$, written in terms of 
$Q^\ast_i:= Q^\ast_{il}\tau_l$ with the Pauli matrices $\tau_i$.
In the strong coupling limit the groundstate wave functional (\ref{PsiQ})
reduces to the very simple form
\be
\label{Psihomo}
\Psi[\phi_1,\phi_2,\phi_3]=\exp\left[-g\phi_1\phi_2\phi_3\right]~.
\ee
It is nonnormalizable despite the sign definiteness of its exponent
($\phi_i > 0~, \ i=1,2,3$).
For the analysis of this wave function in the neighbourhood of the line
$\phi_1=\phi_2=0$ of minima of the classical potential (\ref{Vhomo}),
it is useful to pass from the variables $\phi_1$ and $\phi_2$ transverse to
the valley to the new variables $\phi_\perp$ and $\gamma$ via
\be
\phi_1=\phi_\perp \cos\gamma\ , \ \phi_2=\phi_\perp\sin\gamma
\ \ \left(\phi_\perp\ge 0~,\  0\le\gamma \le {\pi\over 2}\right)~.
\ee
The classical potential then reads
\be
V(\phi_3,\phi_\perp,\gamma)=g^2\left(\phi_3^2\phi_\perp^2+{1\over 4}
\phi_\perp^4\sin^2(2\gamma)\right)~,
\ee
and the groundstate wave function (\ref{Psihomo}) becomes
\be
\Phi[\phi_3,\phi_\perp,\gamma]=
\exp\left[-{1\over 2}g\phi_3\phi_\perp^2\sin(2\gamma)\right]~.
\ee
We see that close to the bottom of the valley, for small $\phi_\perp$, the
potential is that of a harmonic oscillator
and the wave functional correspondingly a Gaussian with a maximum at the
classical minimum line $\phi_\perp=0$. The height of the maximum is constant
along the valley. The non-normalizability of the groundstate wave function 
(\ref{Psihomo}) is therefore due to the outflow of the
wave function with constant values along the valley to arbitrarily
large values of the field $\phi_3$. The formation of
condensates with macroscopically large fluctuations of the field amplitude
might be a very interesting consequence of the properties of the
classical potential.
To establish the connection between this phenomenon and the model of the
squeezed gluon condensate \cite{BlaPa} will be an interesting task for
further investigation.

We now would like to find the effective classical field theory
to which the unconstrained theory reduces in the limit of infinite
coupling constant $g$, if we assume that the classical system spontaneously
chooses one of the classical zero energy minima of the leading order $g^2$ 
part (\ref{Vhomo}) of the potential. 
As discussed above these classical minima include apart from the perturbative 
vacuum, where all fields vanish, also field configurations with one scalar 
field attaining arbitrary values. 
Let us therefore put without loss of generality 
(explicitly breaking the cyclic symmetry)
\be
\phi_1=\phi_2=0\ ,\ \ \ \phi_3\ -\ {\rm arbitrary}~,
\ee
such that the potential (\ref{Vhomo}) vanishes.
In this case the part of the potential (\ref{Vinhom1}) containing 
derivatives takes the form
\bea
V_{\rm inh} &=&
 \phi_3(x)^2\big[(\Gamma^2_{{\ }1 3}(x))^2+(\Gamma^2_{{\ }2 3}(x))^2
           +(\Gamma^2_{{\ }3 3}(x))^2+\nonumber\\
&& \ \ \ \ \ \ \ \     +(\Gamma^3_{{\ }1 1}(x))^2+(\Gamma^3_{{\ }2 1}(x))^2
           +(\Gamma^3_{{\ }3 1}(x))^2 \big]+\nonumber\\
&&  +2\phi_3(x)\big[\Gamma_{{\ }3 1}^3(x) X_1\phi_3
                   +\Gamma_{{\ }3 2}^3(x) X_2\phi_3\big]+\nonumber\\
&& +\big[(X_1\phi_3)^2+(X_2\phi_3)^2\big]~.
\eea
Introducing the unit vector
\be
n_i(\phi,\theta):=R_{3i}(\phi,\theta)~,
\ee
pointing along the 3-axis of the ``intrinsic frame'', one can write
\bea
V_{\rm inh} &=& \phi_3(x)^2 \left(\partial_i {\vec n}\right)^2
              +(\partial_i\phi_3)^2
 -(n_i \partial_i\phi_3)^2 \nonumber\\ 
&& \ \ \ \ \ \ \ \ - (n_i\partial_i n_j) \partial_j (\phi_3^2)~.
\eea
Concerning the contribution from the nonlocal term in this phase,
we obtain for the leading part of the electric fields
\be
{\cal E}^{(0)}_1= -\xi_1/\phi_3\ \ ,\ \ {\cal E}^{(0)}_2=-\xi_2/\phi_3~.
\ee
Since the third component ${\cal E}^{(0)}_3$ and ${\cal P}_3$ are singular
in the limit
$\phi_1,\phi_2\rightarrow 0$, it is necessary to have $\xi_3\rightarrow 0$.
The assumption of a definite value of $\xi_3$ is in accordance with the fact
that the potential is symmetric around the 3-axis for small $\phi_1$ and
$\phi_2$, such that the intrinsic angular momentum $\xi_3$
is conserved in the neighbourhood of this configuration.
Hence we obtain the following effective Hamiltonian up to  order $O(1/g)$
\bea
H_{\rm eff} &=& {1\over 2}\int d^3x
\bigg[\pi_3^2+{1\over \phi_3^2}(\xi_1^2+\xi_2^2) +(\partial_i\phi_3)^2
+\phi_3^2(\partial_i{\vec n})^2 \nonumber\\
&&\ \ \ \ \ \ \ \ \ \ \ \ \ \ \
 -(n_i \partial_i\phi_3)^2 - (n_i\partial_i n_j) \partial_j (\phi_3^2)
\bigg]~.
\eea
After the inverse Lagrangian transformation we obtain the corresponding
nonlinear sigma model type effective Lagrangian for  the unit vector
${\vec n}(t,{\vec x})$ coupled to the scalar field $\phi_3(t,\vec{x})$
\bea
\label{Leff}
L_{\rm eff}[\phi_3,{\vec n}]&=& {1\over 2}\int d^3x 
\bigg[(\partial_\mu \phi_3^2)^2+
       \phi_3^2(\partial_\mu {\vec n})^2\nonumber\\
&& \ \ \ \ \ \       +(n_i \partial_i\phi_3)^2 
      + n_i(\partial_i n_j) \partial_j (\phi_3^2)\bigg]~.
\eea
In the limit of infinite coupling the unconstrained field theory in terms of
six physical fields equivalent to the original $SU(2)$ Yang-Mills theory 
in terms of the gauge fields $A_\mu^a$ reduces therefore to an effective 
classical field theory involving only one of the three scalar fields and
two of the three rotational fields summarized in the unit vector $\vec{n}$.  
Note that this nonlinear sigma model type Lagrangian
admits singular hedgehog configurations of the unit vector field $\vec{n}$.
Due to the absence of a scale at the classical level, however, these are
unstable. 
Consider for example the case of one static monopole placed at the origin,
\be
n_i:= x_i/r~,\ \ \ \phi_3=\phi_3(r)~, \ \ \ r:=\sqrt{x_1^2+x_2^2+x_3^2}~.
\ee
Minimizing its total energy $E$ 
\be
E[\phi_3]=4\pi \int dr \phi_3^2(r)
\ee 
with respect to $\phi_3(r)$ we find the classical solution $\phi_3(r)\equiv 0$.
There is no scale in the classical theory.
Only in a quantum investigation a mass scale such as a nonvanishing value for
the condensate $<0|\hat{\phi}_3^2|0>$ may appear, which might be related to 
the string tension of flux tubes directed along the unit-vector field 
${\vec n}(t,{\vec x})$. 
The singular hedgehog configurations of such string-like directed flux tubes 
might then be associated with the glueballs.
The pure quantum object $<0|\hat{\phi}_3^2|0>$ might be realized as a squeezed 
gluon condensate \cite{BlaPa}.
Note that for the case of a spatially constant condensate,
\be
<0|\hat{\phi}_3^2|0>=:2 m^2= const.~,
\ee 
the quantum effective action corresponding to (\ref{Leff}) should 
reduce to the lowest order term of the effective soliton Lagangian discussed
very recently by Faddeev and Niemi \cite{FadNiem}
\be
\label{FN}
L_{\rm eff}[{\vec n}]= m^2\int d^3x (\partial_\mu {\vec n})^2~.
\ee
As discussed in \cite{FadNiem}, for the stability of these knots 
furthermore a higher order Skyrmion-like term in the derivative expansion 
of the unit-vector field ${\vec n}(t,{\vec x})$ is necessary.
To obtain it from the corresponding higher order terms in the strong coupling 
expansion of the unconstrained Hamiltonian (\ref{eq:unchdia}) is under present 
investigation.

In summary we have found a representation
of the physical variables  which is appropriate for the
study of the infrared limit of $SU(2)$ Yang-Mills theory.   
We have shown how in the infrared limit an effective nonlinear sigma model 
type Lagrangian can be derived which out of the six physical fields involves 
only one of three scalar fields and two rotational fields summarized 
in a unit vector. The study of the corresponding quantum theory as well as 
the consideration of higher order terms in the strong coupling expansion
will be the subject of future work.

\bigskip

\noindent
{\bf Acknowledgements}
\bigskip

We are grateful for discussions with  S.A. Gogilidze, D.M. Mladenov,
V.N. Pervushin, G. R\"opke, A.N. Tavkhelidze and J.Wambach.
One of us (A.K.)  acknowledges the Deutsche Forschungsgemeinschaft for
providing a visting stipend.
His work was also supported  by the Russian Foundation for
Basic Research under grant No. 96-01-01223.
H.-P. P. acknowledges support by the Deutsche Forschungsgemeinschaft
under grant No. Ro 905/11-2 and by the Heisenberg-Landau program .


\end{document}